# Optical polarization of nuclear ensembles in diamond

Ran Fischer[1,*], Andrey Jarmola[2,3], Pauli Kehayias[2], and Dmitry Budker[2,4,+]

1. Department of Physics, Technion – Israel Institute of Technology, Haifa 32000, Israel
2. Department of Physics, University of California, Berkeley, CA 94720-7300, USA
3. Laser Centre, The University of Latvia, Rainis Boulevard 19, 1586 Riga, Latvia
4. Nuclear Science Division, Lawrence Berkeley National Laboratory, Berkeley, CA 94720, USA

We report polarization of a dense nuclear-spin ensemble in diamond and its dependence on magnetic field and temperature. The polarization method is based on the transfer of electron spin polarization of negatively charged nitrogen vacancy color centers to the nuclear spins via the excited-state level anti-crossing of the center. We polarize 90% of the $^{14}$N nuclear spins within the NV centers, and 70% of the proximal $^{13}$C nuclear spins with hyperfine interaction strength of 13-14 MHz. Magnetic-field dependence of the polarization reveals sharp decrease in polarization at specific field values corresponding to cross-relaxation with substitutional nitrogen centers, while temperature dependence of the polarization reveals that high polarization persists down to 50 K. This work enables polarization of the $^{13}$C in bulk diamond, which is of interest in applications of nuclear magnetic resonance, in quantum memories of hybrid quantum devices, and in sensing.

## I. INTRODUCTION

Nuclear spins are a natural choice for applications requiring long relaxation times due to their immunity to unwanted perturbations from the environment. Among the many applications are magnetic resonance-based bio-sensing [1], rotation sensing [2-4], and quantum computing [5-7]. Specifically, $^{13}$C nuclear spins in diamond weakly interact with the lattice, resulting in long relaxation times, up to several hours [8]. These relaxation rates mark nuclear spins in diamond as ideal for the applications. However, thermal polarization of nuclear spins is limited by the Boltzmann factor, which close to unity, especially at room temperature. This limitation results in poor sensitivities, requiring methods that achieve an enhancement over thermal polarization. Dynamic nuclear polarization (DNP) allows us to achieve higher

Earlier work on DNP in diamond included ensemble polarization at cryogenic temperatures [9–12] as well as single-spin polarization at room temperature [13, 14]. Here we demonstrate a high polarization degree of a nuclear ensemble at room temperature. This is beneficial for obtaining high signal-to-noise ratio in precision measurements based on nuclear spins and in initializing and manipulating collective quantum memories [15]. Based on the early results of this work (posted on the arXiv [16]), bulk polarization of $^{13}$C in diamond at room temperature was achieved by exploiting the optical polarization of the negatively charged nitrogen vacancy (NV) color center [17]. This study sets the basis for bulk polarization, while illuminating the role of the temperature in achieving a high polarization degree.

Here, we demonstrate polarization of a dense ensemble of nuclear spins in diamond, which comprises approximately $10^{10}$ spins, by applying the method of Ref. [13], demonstrated for single $^{15}$N, $^{14}$N, and $^{13}$C nuclear spins [13, 14] . This method is based on the transfer of the polarization of the electron spins of NV color centers to the nuclear spins at the excited-state level anti-crossing (ESLAC) of the center. In Refs. [13, 14], the method was demonstrated with single centers, while its applicability to ensembles is non-trivial due to the local effects such as strain. We show that averaging of the orbital angular momentum by phonon interactions [18] enables the ESLAC polarization technique even for dense ensembles of NV centers. The temperature dependence of the polarization process is studied, displaying the transition from total quenching of the orbital momentum at room temperature to the presence of spin-orbit splitting at cryogenic temperatures, resulting in a failure of the polarization process. The latter is accompanied by excited-state optically detected magnetic resonance (ESODMR) spectrum measurements in an attempt to correlate the energy-level structure of the excited-state with the nuclear polarization.

We demonstrate a polarization of 90% for the $^{14}$N nuclear spins within the NV centers, as well as a 70% polarization of proximal $^{13}$C nuclear spins. We study polarization dependence on the magnetic-field magnitude, showing a similar behavior as in the single-center study [13], but with an additional effect of the ground-state cross-relaxation of the NV centers with the substitutional-nitrogen (P1) centers. An attempt to reproduce the experimental results by a numerical model accompanies both the magnetic field and the temperature data. The model is based on the steady-state Lindblad equation for the density matrix. Further details on the model are presented in the Appendix.

The NV center in diamond is comprised of a substitutional nitrogen atom and a vacancy at an adjacent lattice site. It has spin-triplet ground $^3A_2$ and excited $^3E$ states connected with an optical transition [19]. The ground state has a zero-field splitting between the $m_s = 0$ and the $m_s = \pm 1$ spin sublevels of 2.87 GHz. The $^3A_2 \leftrightarrow {}^3E$ transition has a zero-phonon line at 638 nm, and a phonon side-band that spans from 638 nm to about 460 nm in absorption [20]. Optical excitation of the NV center results in optical pumping into the $m_s = 0$ sublevel due to an intersystem crossing through the singlet levels that preferentially transfers the $m_s = \pm 1$ states from the $^3E$ excited-state to the $m_s = 0$ sublevel in the ground $^3A_2$ state. Moreover, the intersystem crossing results in a higher fluorescence of the $m_s = 0$ state than the $m_s = \pm 1$ states upon $^3A_2 \to {}^3E$ excitation, enabling optical spin read-out and detection of the ground-state spin transitions by optically detected magnetic resonance (ODMR) [21]. Through hyperfine interaction, ODMR can also detect nuclear spins in diamond such as those of $^{13}$C, $^{14}$N, and $^{15}$N.

## II. METHOD

The polarization process can be understood by examining the $^3E$ excited-state Hamiltonian of the NV center (Fig. 1, where, for illustration, we ignore the nitrogen nuclear spin but do include nuclear spin of a proximal $^{13}$C). At room temperature, the spin-orbit terms and strain terms are averaged out to zero [15], thus the excited-state Hamiltonian can be written as [22]:

$$H_{es} = D_{es}\left(S_z^2 - \frac{1}{3}\vec{S}^2\right) + E_{es}\left(S_X^2 - S_Y^2\right) + \vec{B}\cdot\left(\gamma_{NV}\vec{S} + \gamma_N\vec{I}\right) + \vec{I}\cdot A\cdot\vec{S}. \quad (1)$$

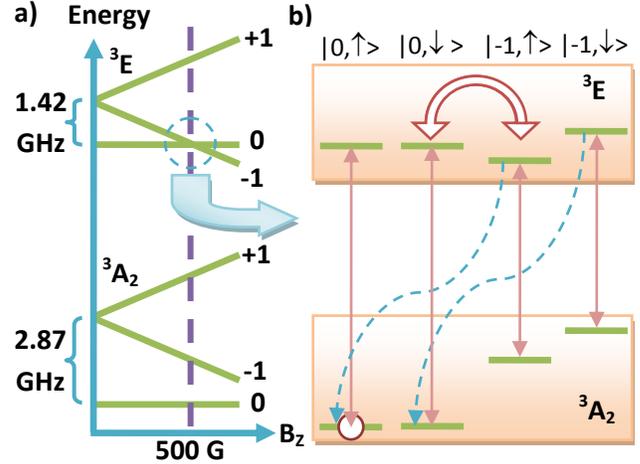

Fig. 1 - (color online) Nuclear-spin polarization process of the $^{13}$C nuclear spin. The $^3A_2$ and the $^3E$ are the triplet ground and excited states respectively. a) Energy diagram of the electronic triplet levels of the NV center at room temperature as a function of the axial magnetic field. The ESLAC of the spin-states $m_S = 0$ and $m_S = -1$ can be seen in a field of ≈ 500 G. b) The joint manifold of the nuclear spin of the $^{13}$C and the electronic spin of the NV center near the ESLAC. The states are labeled as $|m_S, m_I\rangle$. At 500 G, the electronic sub-states are nearly degenerate in the excited-state, while in the ground-state there is a splitting of ~1.5 GHz. The axial hyperfine interaction can be seen as a splitting of the $m_S=-1$ states, while the off-axial hyperfine interaction mixing the electronic and the nuclear spins is represented by the bi-directional curved arrow in the $^3E$ level. The intersystem crossing is represented by dashed arrows connecting the $m_S = -1$ in the excited state to the $m_S = 0$ in the ground-state. The optical transitions are represented by bi-directional straight arrows, which conserve spin.

The first term gives rise to the zero-field splitting that separates the electron-spin sub-states $m_s = 0$ and $m_s = \pm 1$ by $D_{es}$; the second term is the spin-strain interaction, taken along the x-axis, which reduces the symmetry and lifts the degeneracy at zero magnetic field of the $|\pm 1\rangle$ states. The third term is the interaction of the electron and nuclear spins with an external magnetic field. Here $\gamma_{NV}$ is the gyromagnetic ratio of the NV center and $\gamma_N$ is the gyromagnetic ratio of the nuclear spin, typically, three orders of magnitude smaller than $\gamma_{NV}$. The last term is the hyperfine interaction of the nuclear and electron spins of the NV center, where $A$ is the hyperfine-interaction tensor. The spin-strain interaction $E_{es}\left(S_X^2 - S_Y^2\right)$ is generally negligible for an axial magnetic field near the ESLAC, even for a highly doped type-Ib diamond. Moreover, if the off-axial magnetic-field term is small compared to the excited-state natural linewidth, it can be neglected. This leaves the hyperfine interaction the only non-secular term in the Hamiltonian. For the nitrogen nucleus associated with the NV center, the interaction is along the NV center's axis, and is of the form of:

$$H_{HF} = \vec{I} \cdot A \cdot \vec{S} = A_{par} I_z S_z + \frac{A_{perp}}{2}(I_+ S_- + I_- S_+), \quad (2)$$

where the magnitudes of coefficients $D_{es}, A_{par}, A_{perp}$ are: $D_{es} \approx 1.4$ GHz, and $A_{perp}, A_{par} \approx$ tens of MHz [22]. For a magnetic field that corresponds to $B_z \approx D_{es}/\gamma_{NV} \approx 500$ G, the diagonal terms of $m_s = -1$ approach the diagonal terms of $m_s = 0$, resulting in a mixing of the $|m_s = 0, m_I = m\rangle$ and $|m_s = -1, m_I = m+1\rangle$ states due to the off-axial hyperfine interaction. The off-axial hyperfine interaction $A_{perp}$ is on the order of tens of MHz, which is greater than the excited-state decay rate. As a result, the mixing in the excited state takes place before the state decays. Since the electron spin of the NV center is optically pumped to $m_s = 0$ and the $|m_s = 0, m_I = I\rangle$ state is not mixed by the hyperfine interaction, the nuclear spin is pumped toward its maximum projection. It should be noted that in the general case, when the interaction is not along the NV center's axis, the hyperfine tensor A is anisotropic. This affects the degree and orientation of polarization that can be reached with this method. Therefore, different nuclear species, such as $^{14}$N and $^{13}$C, as well as spins in different positions may exhibit different degrees of polarization [23]. Furthermore, the anisotropic hyperfine interaction can lead to positive and negative polarization, enabling polarization control [17, 23].

An essential process that enables nuclear polarization through the ESLAC is quenching of the orbital momentum through phonon interaction. At low temperatures, the spin-orbit and strain-orbit terms dominate the energy-level structure of the excited-state. Specifically, in a type-Ib diamond, strain can be large and dependent on location of the center in the sample. This locally splits the excited state, resulting in disappearance of the anti-crossing at 500 G. At room temperature, the quenching of the orbital momentum averages the spin-orbit and strain splittings to zero. The impact of this quenching process was studied in emission [18] and ground-state ODMR [25]. Indeed, at room temperature, ESLAC is observed near 500 G, even in type-Ib diamonds, thus enabling the nuclear-polarization process.

### III. EXPERIMENTAL SETUP

We used a type-Ib diamond with 50 ppm of initial nitrogen concentration that was irradiated with 3 MeV electrons and annealed at 750°C for 2 hours. The NV-center concentration after this process is estimated to be ∼10 ppm. The inhomogeneously broadened full width half maximum of the ground-state resonance is ∼2 MHz, so the hyperfine splitting in the ground state (2.16 MHz) due to interaction with $^{14}$N is resolved [25, 26].

We employed a conventional ODMR setup with a focusing lens used for excitation with a 532 nm laser and for fluorescence collection. Microwave radiation is applied with a copper wire placed on top of the diamond near the excitation spot. The diamond is placed inside a liquid-helium flow cryostat (Janis ST-500), with a sample extension for magnetic measurements. The cryostat was equipped with a heater, enabling the control of the temperature from liquid-helium temperatures to above the room temperature, using a temperature controller (Lakeshore 331). The magnetic field was applied either with a permanent magnet, or using a solenoid. In the latter case, the maximum applied field was 550 G. The magnetic field was aligned to better than 1° from the [111] axis of the diamond crystal with a tilt stage. The alignment of the magnetic field with respect to one of the NV centers' orientations is crucial for achieving a high degree of nuclear spin polarization [13]. Note that the polarization process requires only optical excitation that pumps the electronic spin of the NV center and a proper alignment and magnitude of the magnetic field. It does not require an application of MW or radio-frequency radiation as in conventional DNP methods.

### IV. RESULTS AND DISCUSSION

Figure 2 shows a comparison of the ODMR signal contrast for the [111] orientation at low magnetic field and at the ESLAC. The contrast is defined as the fractional change in fluorescence when the microwaves are applied. The ODMR signal at low magnetic field, the top trace in Fig. 2, exhibits equal contrast on the three hyperfine components of the resonance (unpolarized $^{14}$N nuclei), while the ODMR signal near ESLAC (bottom trace in Fig. 2) exhibits only one dominant resonance with approximately three times the contrast of that at low field. We characterize the nuclear spin polarization by analyzing the contrast of ODMR signals, calculating the polarization as:

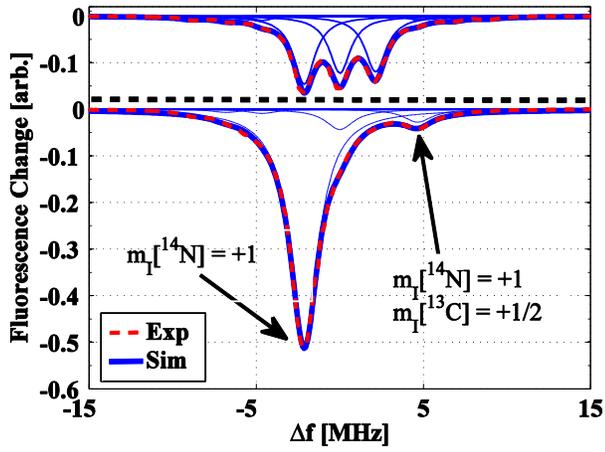
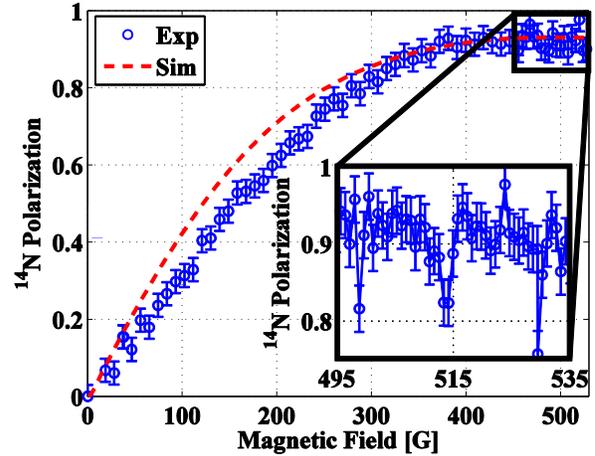

Fig. 2 – (color online) Comparison of ODMR signal of the [111] orientation in a low magnetic field and near the ESLAC. Top figure: ODMR signal at 21 G. Bottom figure: ODMR signal near the ESLAC at 387 G. Thick full curve correspond to the experimental results, thick dashed curve correspond to the fitted multiple resonances. Narrow curves correspond to the fitted individual resonances. The high peak marked with an arrow corresponds to the ($m_S = 0$, $^{14N}m_I = +1 \rightarrow m_S = +1$, $^{14N}m_I = +1$) transition. The marked low peak corresponds to ($m_S = 0$, $m_I[^{14}N] = +1$, $m_I[^{13}C] = +1/2 \rightarrow m_S = +1$, $m_I[^{14}N] = +1$, $^{13C}m_I[^{13}C] = +1/2$) transition, and it is due to hyperfine interaction with $^{14}N$ and $^{13}C$ nuclear spins. Top figure: $f_{resonance}$ = 2.93 GHz. Bottom figure: $f_{resonance}$ = 3.92 GHz.

Fig. 3 - (color online) Room-temperature 14N nuclear polarization as a function of the axial magnetic field. The circles are the experimental measurements, and the dashed curve is the simulation result at zero strain. Inset: $^{14}N$ polarization in a smaller magnetic field range showing three dips of the nuclear polarization corresponding to cross-relaxation with P1 centers.

$$P = \frac{\sum_i m_i N_i(m_i)}{I \sum_i A_i(m_i)}, \quad (3)$$

where $N_i$ is the relative population of the sub-state $m_i$, and $I$ is the nuclear spin. The relative population was estimated as the ratio of the ODMR-resonance amplitude to the sum of amplitudes of all nuclear sub-states. We estimate that the polarization of the $^{14}N$ nuclear spins that are a part of the NV centers is 92±3%.

Since we deal here with an ensemble of NV centers, some centers have a proximal $^{13}C$ nucleus that interacts with the NV center. These NV centers exhibit a double hyperfine splitting – one due to the $^{14}N$ nucleus, and one due to the $^{13}C$ nucleus [23]. The $^{14}N$ hyperfine constant ($A_{par}$) is 2.16 MHz in the NV ground state [26], while the value for an interaction with $^{13}C$ depends on its position with respect to the NV center. A $^{13}C$ located in the lattice sites nearest to the vacancy gives a splitting between the hyperfine side-resonances in the ODMR spectrum of 130 MHz, while the splitting for the next closest $^{13}C$ spins in is 13.8 MHz and 12.8 MHz [23, 28]. In a diamond with a natural abundance of $^{13}C$ each NV center has 1.1% percent probability to have a $^{13}C$ nucleus in each of the relevant positions, while the trigonal symmetry lead to a product of 1.1% and 3 or 6 equivalent positions. Moreover, since the ground-state resonance width is ~2 MHz, only interactions with hyperfine splitting greater than ~10 MHz are clearly resolved (130 MHz, 13.8 MHz, 12.8 MHz). The central resonance, not-split by $^{13}C$ hyperfine interaction dominates the ODMR signal. Resonances split by $^{13}C$ interaction with a splitting smaller than 10 MHz, corresponding to an offset of 5 MHz, are obscured by the central resonance, and are consequently unresolved.

In an ensemble of NV centers aligned along the magnetic field, the non-axial hyperfine interaction at the ESLAC can results in polarization of both nuclear-spin species [23]. This means that the proximal $^{13}C$ nuclear spins are polarized, as well as the $^{14}N$ associated with the NV center. As a result, the ODMR signal of polarized ensemble of nuclear spins of $^{14}N$ and proximal $^{13}C$, which can be seen in the bottom trace of Fig. 2, is composed of a high-contrast resonance that corresponds to polarization of the $^{14}N$ nuclear spin, and a low-contrast resonance (with a contrast of a few percent of that at the primary polarized $^{14}N$ resonance). The low-contrast resonance corresponds to polarization of both $^{14}N$ and the proximal $^{13}C$ nuclear spin, and has a frequency offset from the primary resonance equal to roughly half of the $^{13}C$ hyperfine splitting. We note that the $^{14}N$ hyperfine interaction [23, 27] has an opposite sign relative to the $^{13}C$ interaction, whose strength is 13-14 MHz [23]. We estimate that, in this experiment, a polarization of 70±10% of the resolved proximal $^{13}C$ nuclear spins was reached.

The polarization of the proximal $^{13}$C spins can result in polarization of the bulk $^{13}$C by spin diffusion [12]. In Ref. [12] it was estimated that, for a diamond with NV concentration of ~ 10 ppm with a natural abundance of $^{13}$C, spin diffusion should polarize the bulk $^{13}$C which are found in the optically excited spot in a typical time of tens of seconds. This requires that the bulk $^{13}$C should have a longitudinal spin-relaxation time at room temperature of about a minute. The room-temperature longitudinal spin-relaxation time of $^{13}$C in a type-Ib diamond with 10 ppm of nitrogen was measured in Ref. [8] to be more than 1 hour, however in a multi-tesla magnetic field.

The measurement of the magnetic-field dependence of the degree of polarization has revealed a difference between nuclear polarization in an ensemble in a N-rich diamond and a single center in a N-dilute diamond, [13]. The general form of the dependence is similar; however, the degree of polarization decreases by ~15% at 499 G, 514 G and 529 G. These dips correspond to cross-relaxation with P1 centers, with its hyperfine structure, which reduces the ground-state electron-spin polarization [29], and as a result, reduces the nuclear polarization. The general form of the experimental results appearing in Fig. 3 is reproduced by the aforementioned numerical model, for zero-strain. From the similarity of the polarization degree of the single center study and our results on a dense ensemble, and the reproduction of the results by the simulation at zero-strain, we conclude that the efficiency of the nuclear polarization process is weakly affected by strain, which can be considerable for diamonds with abundant P1 centers. Furthermore, the high nitrogen concentration results in cross-relaxation dips in the polarization curve.

The nuclear polarization process is enabled at room temperature by quenching of the orbital momentum of the excited-state of the NV center, leading to a three sub-state energy-level structure depicted in Fig. 1a. Understanding the transition from cryogenic temperatures to room temperature can illuminate the role of the phononic interaction in the excited-state and its impact on the nuclear polarization at the ESLAC. To this end, we studied the temperature dependence of nuclear polarization at the ESLAC, accompanied by a zero-field ESODMR measurement in a range of temperatures from 10 to 300 K, Fig. 4(a, b). The nuclear polarization shows a steep increase from zero polarization at 20 K to ~0.6 at 50 K, and a further gradual increase with increasing temperature above 50 K till leveling-off at a high level above 200 K. The ESODMR at zero-field was measured by sweeping the MW frequency in the 1-1.8 GHz range. At zero field the ground-state transitions occur at ~2.87 GHz, suggesting that the resonances at the 1-1.8 GHz correspond to the excited-state structure, originating from the zero-field splitting $D_{ES}$ and the inhomogeneous spin-strain interaction $E_{ES}$. Fig. 4b depicts a product of the average over the central range of the resonance in the range of 1.38 - 1.44 GHz and the full-width at half maximum (FWHM) as a function of temperature. This parameter represents the integral over the resonance. The behavior of the contrast reveals a similar pattern to that of $^{14}$N nuclear polarization - a steep increase from nearly zero contrast between 40-60 K, and a gradual increase from 60 K to 300 K. The ESODMR lineshape reveals broadening of the resonance with decreasing temperature, see inset in Fig. 4b, while at low temperatures the ESODMR resonance vanishes completely. Interestingly, temperature dependence of the ESODMR signal and the $^{14}$N nuclear polarization are similar but not identical. Specifically, the ESODMR nearly vanishes below 50 K, while the nuclear polarization persists down to 20 K, at a level of tens of percents. To study the relation between the ESODMR line-width and the $^{14}$N polarization we assume that when the temperature decreases the orbit-strain interaction comes into play due to a decrease in the efficiency in the quenching of the orbital momentum. Under this assumption we use the numerical model, introduced in section II, to infer the $^{14}$N polarization from strain distribution as a function of temperature extracted from the ESODMR spectrum, Fig. 4b. For simplicity, we neglected all the terms in the Hamiltonian aside from the strain terms, while the latter were referred as a Gaussian distribution, which was fitted to the zero-field ESODMR spectrum, seen in Fig. 4b. The result of the model is given in Fig. 4a, exhibiting a general agreement with the experimental data. Finally, both the nuclear polarization and the ESODMR appear to be controlled by two processes which come into play at a different temperature range.

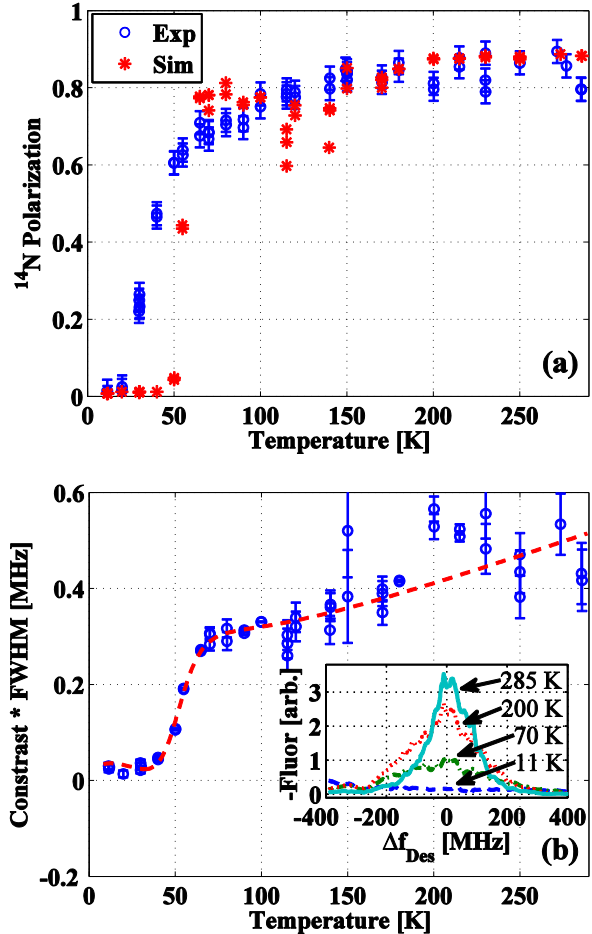

Fig. 4 - (color online) (a) Temperature dependence of $^{14}$N polarization. Circles represent experimental results; asterisks represent simulated $^{14}$N polarization, in which the strain distribution is extracted from the ESODMR measurements. (b) Temperature dependence of zero-field excited-state ODMR. The main figure represents a product of the average over the central range of the resonance in the range of 1.38 - 1.44 GHz, and the full-width half maximum. The dashed curve is a guide for an eye. Inset: Excited-state ODMR at zero-field for four different temperatures between 10 K and 285 K.

Understanding the nature of these processes, and more accurate modeling of the $^{14}$N polarization and the ESODMR dependence on temperature requires further study. The ESODMR results contrast a previous study on a single center [21], in which the ESODMR persists only down to 150 K. We suspect that the difference originates in the nature of the present measurements, which include an ensemble of NV centers with broad distribution of strain, rather than a single strain value. Moreover, a different study, that monitored the fluorescence versus magnetic field, and was performed on a dense ensemble of NV centers showed a similar behavior, where the ESLAC remained visible down to 40 K [18].

## V. CONCLUSIONS AND OUTLOOK

We have demonstrated a high polarization degree of a dense ensemble of nuclear spins associated with NV centers in a type-Ib diamond at room temperature. The polarization method is based on the hyperfine interaction at the ESLAC, and was applied to proximal $^{13}$C as well as the $^{14}$N nuclear spins within the centers. We estimate that a polarization degree of 70±10% of the resolved proximal $^{13}$C nuclear spins and 92±3% of the $^{14}$N nuclear spins was achieved. The polarization degree achieved here is similar to earlier single-center studies [13, 14].

The dependence of the degree of polarization of the $^{14}$N nuclear spins on the magnetic field was studied, displaying similar pattern to previous studies on single centers in low-doping samples. However, cross-relaxation of the NV center with substitutional nitrogen in the vicinity of the ESLAC decreases the center's electron polarization, and thus, the nuclear-spin polarization as well.

The temperature dependence of nuclear polarization was measured, revealing a high polarization degree above 50 K. We accompanied the polarization measurement with recording zero-field ESODMR spectra at the same temperatures, showing features in the temperature dependence that correlate with those for the nuclear-polarization temperature dependence. The results reveal the robustness of the nuclear polarization process through the ESLAC, where a high nuclear polarization is reached even at relatively low temperatures, although the phonon density is considerably reduced. Furthermore, this result implies that the excited-state energy structure is dominated by the spin-spin interaction rather than by the spin-orbit interaction above 50 K. A density-matrix model was developed to reproduce the temperature dependence of the $^{14}$N polarization, reconstructing the spin-strain interaction from the ESODMR spectrum. This temperature study of the excited state can open a path for understanding the transition from the low-temperature regime to room temperatures through phonon-orbit interaction - a regime which is currently not well understood.

We envision the following applications of this work. First, the polarization of the proximal $^{13}$C nuclear spins enables polarization, by spin diffusion, of the $^{13}$C in the bulk of the diamond crystal [16], enabling applications, for example, in nuclear magnetic resonance and rotation sensing [4, 30]. Bulk polarization may be especially high for $^{13}$C enriched diamonds, which have strong dipolar coupling between neighboring $^{13}$C nuclear spins, compared to natural diamond. Second, for a diamond with low nitrogen content, a high polarization degree of the proximal $^{13}$C nuclear spins may increase the coherence time of the ground state, mainly caused by randomly polarized $^{13}$C nuclear spins.

## ACKNOWLEDGMENTS

The authors thank M. Ledbetter, V. Acosta, N. Manson, P. London, P. Hemmer, M. Doherty and D. English for useful discussions. This work has been supported in part by the AFOSR/DARPA QuASAR program, NSF, IMOD, and by the NATO SFP program. D.B. is grateful for the support by the Miller Institute for Basic Research in Science. A.J. acknowledges support from the ERAF project No.2010/0242/2DP/2.1.1.1.0/10/APIA/VIAA/036. P.K. acknowledges support from the DOE SCGF.

## APPENDIX: DENSITY MATRIX MODEL

To estimate the impact of strain in a type-Ib diamond on the nuclear polarization in the ESLAC, we studied the nuclear polarization dependence on the axial magnetic field and on strain using a numerical model. The numerical model takes into account the triplet excited-state $^3E$ spin states, along with the nuclear spin of the $^{14}$N. The system is modeled by means of solving the steady-state Lindblad equation,

$$\frac{\partial}{\partial t}\rho = -\frac{i}{\hbar}[H,\rho] + L\rho = 0, \qquad (3)$$

where $\rho$ is the density matrix of the NV center $^3E$ triplet and the $^{14}$N nuclear spin, $H$ is the Hamiltonian of Eq. (1), and $L$ is the Lindblad super-operator which effectively describes the decay processes of the NV center and the $^{14}$N nuclear spin. The decay processes include thermal relaxation of both spin species; their rates were taken from the literature [19, 31]. The pumping process was modeled as an asymmetric decay process preferentially populating the $m_s = 0$ spin-state of the NV center. The imbalance in the decay terms of the pumping process was calibrated to fit the typical electronic polarization of the NV center [19]. We note that in this model we didn't consider the $^3A_2$ ground state since the optical transitions $^3A_2 \leftrightarrow ^3E$ are mostly spin preserving, while the pumping process was taken to be a decay-process. Consequently, the model fails to correctly describe the short lifetime of the excited-state, which generally prohibits slow processes. However, here we deal mainly with large terms in the Hamiltonian, substantially faster than the $^3E$ lifetime, and therefore this should not be a major source of inaccuracy.

In addition to the reproduction of the experimental results, seen in Fig. 3 and Fig. 4a, we performed a two dimensional scan (in strain and magnetic field magnitude) of the $^{14}$N nuclear polarization, revealing a weak dependence on both parameters, Fig. 5. Specifically, near the ESLAC at 500 G, a high polarization degree persists for a broad range of axial magnetic fields and strain values. Consequently, we expect a high degree of nuclear polarization for NV centers with strain-spin interaction of up to a few hundreds of MHz.

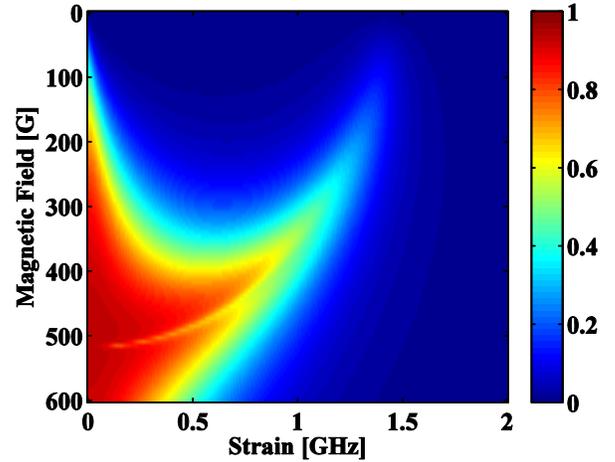

fig. 5 - (color online) A steady-state density matrix simulation of the $^{14}$N nuclear polarization as a function of the axial magnetic field and the spin-strain interaction.

Using this numerical model to describe the temperature dependence of the nuclear polarization we suggest that broad ESODMR spectrum represents the increase of the spin-strain interaction with decreasing temperature. Although we believe that this is a simplification of the physical process, it nevertheless represents a reduction of the center's symmetry due to strain terms in the Hamiltonian, which become dominant at low temperatures.

\* rfischer@techunix.technion.ac.il

+ budker@berkeley.edu

[1] L. Schroder, T. J. Lowery, C. Hilty, D. E. Wemmer, et al., "Molecular Imaging Using a Targeted Magnetic Resonance Hyperpolarized Biosensor", Science Vol. 314, 446 (2006).
[2] K. F. Woodman, P. W. Franks and M. D. Richards, "The Nuclear Magnetic Resonance Gyroscope: a Review", Journal of Navigation 40, 366 (1987).
[3] T. W. Kornack, R. K. Ghosh, M. V. Romalis, "Nuclear Spin Gyroscope Based on an Atomic Comagnetometer", Phys. Rev. Lett. 95, 230801 (2005).
[4] M. P. Ledbetter, K. Jensen, R. Fischer, A. Jarmola, D. Budker, "Gyroscopes based on Nitrogen Vacancy Centers in Diamond", Phys. Rev. A 86, 052116 (2012).
[5] L. M. K. Vandersypen, I. L. Chuang, "NMR Techniques for Quantum Control and Computation", Rev. Mod. Phys. 76, 1037 (2004).
[6] J. J. L. Morton Alexei M. Tyryshkin, Richard M. Brown, Shyam Shankar et al., "Solid-State Quantum Memory Using the 31P Nuclear Spin", Nature Vol. 455, 1085 (2008).
[7] M. V. Gurudev Dutt, L Childress, L Jiang, E Togan et al.., "Quantum Register Based on Individual Electronic and Nuclear Spin Qubits in Diamond", Science Vol. 316, 1312 (2007).
[8] C. J. Terblanche, E. C. Reynhardt, J. A. van Wyk, "13C Spin-Lattice Relaxation in Natural Diamond: Zeeman Relaxtion at

4.7 T and 300 K Due to Fixed Paramagnetic Nitrogen Defects", Solid State Nucl. Magn. Reson. 20, 1 (2001).

[9] E. C. Reynhardt and G. L. High, "Dynamic Nuclear Polarization of Diamond. I. Solid State and Thermal Mixing Effects", J. Chem Phys. 109, 4090 (1998).

[10] E. C. Reynhardt and G. L. High, "Dynamic Nuclear Polarization of Diamond. II. Nuclear Orientation via Electron Spin-Locking", J. Chem Phys. 109, 4100 (1998).

[11] E. C. Reynhardt and G. L. High, "Dynamic Nuclear Polarization of Diamond. III. Paramagnetic Electron Relaxation Times from Enhanced $^{13}$C Nuclear Magnetic Resonance Signals", J. Chem Phys. 113, 744 (2000).

[12] J. King P. J. Coles, J. A. Reimer, "Optical Polarization of 13C Nuclei in Diamond Through Nitrogen-Vacancy Centers", Phys. Rev. B 81, 073201 (2010).

[13] V. Jacques, P. Neumann, J. Beck, M. Markham et al., "Dynamic Polarization of Single Nuclear Spins by Optical Pumping of Nitrogen-Vacancy Color Centers in Diamond at Room Temperature", Phys. Rev. Lett. 102, 057403 (2009).

[14] B. Smeltzer J. McIntyre, L. Childress, "Robust Control of Individual Nuclear Spins in Diamond", Phys. Rev. A 80, 050302 (2009).

[15] D. R. McCamey, J. van Tol, G. W. Morley, C. Boehme, "Electronic Spin Storage in an Electrically Readable Nuclear Spin Memory with a Lifetime > 100 Seconds", Science Vol. 330, 1652 (2010).

[16] R. Fischer, A. Jarmola, P. Kehayias, D. Budker, "Room-Temperature Optical Polarization of Nuclear Ensembles in Diamond", arXiv:1202.1072v2 (2012).

[17] R. Fischer C. Bretschneider, P. London, D. Budker et al., "Bulk Nuclear Polarization Enhanced at Room-Temperature by Optical Pumping", arXiv:1211.5801 (2012).

[18] L. J. Rogers R. L. McMurtrie, M. J. Sellars and N. B. Manson, "Time-Averaging within the Excited State of the Nitrogen-Vacancy Centre in Diamond", New J. Phys. 11, 063007 (2009).

[19] N. B. Manson, J. P. Harrison, M. J. Sellars, "Nitrogen-Vacancy Center in Diamond: Model of the Electronic Structure and Associated Dynamics", Phys. Rev. B 74, 104303 (2006).

[20] V. M. Acosta, E. Bauch, M. P. Ledbetter, C. Santori et al., "Diamonds with a High Density of Nitrogen-Vacancy Centers for Magnetometry Applications", Phys. Rev. B 80, 115202 (2009).

[21] A. Gruber, A. Dräbenstedt, C. Tietz, L. Fleury et al., "Scanning Confocal Optical Microscopy and Magnetic Resonance in Single Defect Centers", Science Vol. 276, 2012 (1997).

[22] G. D. Fuchs, V. V. Dobrovitski, R. Hanson, A. Batra et al., "Excited-State Spectroscopy Using Single Spin Manipulation in Diamond", Phys. Rev. Lett. 101, 117601 (2008).

[23] B. Smeltzer, L. Childress, A. Gali, "$^{13}$C Hyperfine Interactions in the Nitrogen-Vacancy Centre in Diamond, New J. Phys. 13, 025021 (2011).

[24] H. Wang, C. S. Shin, C. E. Avalos, S. J. Seltzer et al., "Sensitive Magnetic Control of Ensemble Nuclear Spin Hyperpolarization in Diamond (2012), arXiv:1212.0035 (2012).

[25] A. Batalov, V. Jacques, F. Kaiser, P. Siyushev et al., "Low Temperature Studies of the Excited-State Structure of Negatively Charged Nitrogen Vacancy Color Center in Diamond", Phys. Rev. Lett. 102, 195506 (2009).

[26] M. Steiner, P. Neumann, J. Beck, F. Jelezko et al., "Universal Enhancement of the Optical Readout Fidelity of Single Electron Spins at Nitrogen-Vacancy Centers in Diamond", Phys. Rev. B 81, 035205 (2010).

[27] S. Felton, A. M. Edmonds, M. E. Newton, P. M. Martineau et al., "Hyperfine Interaction in the Ground-state of the Negatively Charged Nitrogen Vacancy Center in Diamond", Phys. Rev. B 79, 075203 (2009).

[28] A. Dreau, J. R. Maze, M. Lesik, J. F. Roch et al., "High-Resolution Spectroscopy of Single NV Defects Coupled with Nearby $^{13}$C Nuclear Spins in Diamond", Phys. Rev. B 85 134107 (2012).

[29] S. Armstrong, L. J. Rogers, R. L. McMurtrie, N. B. Manson, "NV-NV Electron-Electron Spin and NV-NS Electron-Electron and Electron-Nuclear Spin Interaction in Diamond", Physics Procedia 3, 1569 (2010).

[30] A. Ajoy, P. Cappellaro, "Stable Three-Axis Nuclear Spin Gyroscope in Diamond", arXiv:1205.1494 (2012).

[31] P. Neumann, J. Beck, M. Steiner, F. Rempp et al, "Single-Shot Readout of a Single Nuclear Spin", Science Vol. 329, 542 (2010).